\documentclass[aps,pra,twocolumn]{revtex4}

\usepackage{amssymb,amsmath,graphicx,graphics,color,epstopdf}

\newcommand{\sech}[1]{\textrm{sech}\left(  #1\right)}

\newcommand{\schr}{Schr\"odinger }

\begin{document}

\title{Constructing new nonlinear evolution equations with supersymmetry}
\author{Rosie Hayward}
\author{Fabio Biancalana}
\affiliation{School of Engineering and Physical Sciences, Heriot-Watt University, EH14 4AS Edinburgh, UK}

\begin{abstract}
The factorisation method commonly used in linear supersymmetric quantum mechanics is extended, such that it can be applied to nonlinear quantum mechanical systems. The new method is distinguishable from the linear formalism, as the superpotential is forced to become eigenfunction-dependent. An example solution is given for the nonlinear \schr equation and its supersymmetric partner equation. This method allows new nonlinear evolution equations to be constructed from the solutions of known nonlinear equations, and has the potential to be a useful tool for mathematicians and physicists working in the field of nonlinear systems, allowing the discovery of previously unknown `dualities' amongst soliton solutions and their respective equations.
\end{abstract}

\maketitle

\section{Introduction}
Supersymmetry (SUSY), a spacetime symmetry, was formulated in the context of quantum field theory as a relationship between fermions and bosons, which provided a means of fixing some crucial problems with the Standard Model of particle physics \cite{Raymond, SUSY}. The concepts of SUSY have since been applied to traditional linear quantum mechanics (QM), creating a new way of understanding relationships between the potentials of the \schr equation \cite{SUSYQM}. Supersymmetry in QM, or SUSY-QM, has had significant impact in the field of optics in particular, where it is instead known as optical SUSY, and has applications in many areas such as mode conversion, transformation optics, and laser arrays \cite{Christo,ChristoN,ChristoT,ChristoC,ChristoL}. However, the focus has not strayed from linear optical problems. Extending this formalism so that it will have potential applications in {\em nonlinear} optics seems a natural way to progress. While there is some previous work on the supersymmetry of the nonlinear \schr equation (NLSE) \cite{SUSYNLS,SUSYNLS2}, it uses the traditional formalism associated with bosons and fermions (by using Grassmann even or odd fields, respectively). Furthermore, research on `nonlinear SUSY' focuses on what is known as N-fold or polynomial SUSY \cite{NLSUSY}, and not on the application of SUSY-QM to nonlinear equations.

In this paper, we show for the first time that the concept of one-dimensional SUSY-QM can be extended to Hamiltonians leading to nonlinear evolution equations, and in particular we present the application of this idea to the NLSE as a representative, important example. It was shown by Bernstein that the NLSE is factorisable and that its eigenstates are tied to the factorisation by a Miura transform \cite{Bernstein}, which, in the context of SUSY-QM, can be realised as the Riccati equation one needs to solve to construct supersymmetric Hamiltonians \cite{SUSYQM}. The traditional method of factorisation is reworked to account for the dependence of a nonlinear Hamiltonian on its own eigenstates.

This method is shown to be applicable to any NLSE-type equation, where the existence of solutions is entirely dependent on whether one can solve the Riccati equation. Furthermore, the `SUSY-partner' equation is only nonlinear if the equation it is derived from is nonlinear, and the superpotential is consequently forced to be eigenstate-dependent. Hence, this formalism is distinct from linear SUSY-QM. Lastly, it is shown that if we consider the nonlinear SUSY partner equation to have an additional level, analogous to the additional ground state level granted to one of a pair of Hamiltonians in linear SUSY-QM, then for this level, the equation reduces to a scale-free non-linear equation, which can be transformed into a linear equation with a simple substitution.

\section{The nonlinear SUSY transformation}
Establishing a SUSY relationship between two Hamiltonians in QM, known as superpartners, relies on being able to factorise the Hamiltonian operator in question \cite{SUSYQM}. For Hamiltonians in linear QM, we can simply require (imposing $\hbar=1$, $m=1/2$):
\begin{equation}
H^{(1)}=-\frac{d^2}{dx^2}+V^{(1)}(x)=\hat{A}^{\dagger}\hat{A}+E_0,
\end{equation}
and,
\begin{equation}
H^{(2)}=-\frac{d^2}{dx^2}+V^{(2)}(x)=\hat{A}\hat{A}^{\dagger}+E_0,
\end{equation}
where $x$ is the spatial variable, $V^{(1,2)}$ are two (in general different) potentials, and $E_0$ is the ground state energy of the first system. The factorisation operators (in a sense analogous to the creation and annihilation operators of the harmonic oscillator) have the form $\hat{A}=d/dx+W(x)$ and $\hat{A}^{\dagger}=-d/dx+W(x)$, where $W(x)$ is known as the {\em superpotential}, a function that connects, and from which one can derive both $V^{(1)}$ and $V^{(2)}$. If the above requirements are met, the two Hamiltonians will share a spectrum of energies denoted by $E_n$, with the exception that an eigenstate corresponding to the ground state energy, $E_0$, will not exist for the system governed by $H^{(2)}$. To transform from one Hamiltonian to the other by means of a SUSY transformation, one must simply solve either of the following Riccati equations:
\begin{equation}
W^2(x)-W'(x)=V^{(1)}(x)-E_0,
\end{equation}
or
\begin{equation}
W^2(x)+W'(x)=V^{(2)}(x)-E_0.
\end{equation}
If the eigenfunctions of $H^{(1)}$ corresponding to energy $E_n$ are denoted by $\phi_n$, and the eigenfunctions of $H^{(2)}$ are denoted $\psi_n$, then they are related by the equations $\psi_n=(E_n-E_0)^{-1/2}\hat{A}\phi_n$ and $\phi_n=(E_n-E_0)^{-1/2}\hat{A}^{\dagger}\psi_n$, when $n\geq1$. In order for the supersymmetry to remain unbroken, the condition $\hat{A}\phi_0=0$ must be upheld, ensuring no eigenfunction corresponding to $E_0$ exists for $H^{(2)}$.

Let us now consider the nonlinear Hamiltonian $H_n=-d^2/dx^2-\kappa|\psi_n|^2$, such than when it acts on $\psi_n$ we recover the stationary NLSE:
\begin{equation}\label{eq:sch}
H_n\psi_n=-\frac{d^2}{dx^2}\psi_n-\kappa|\psi_n|^2\psi_n=E_n\psi_n.
\end{equation}
In order to factorise this Hamiltonian in its general form, we must require our factorisation operators to also be wavefunction-dependent, labelled by the integer $n$. They become $\hat{A}_n=d/dx+W_n(x)$ and $\hat{A}^{\dagger}_n=-d/dx+W_n(x)$, where $W_n(x)$ is a {\em eigenstate-dependent} superpotential. It is important to note at this point that despite the commonly accepted notation, the operator $\hat{A}^{\dagger}_{n}$ is not the Hermitian adjoint of $\hat{A}_{n}$, unlike the operators appearing in the harmonic oscillator problem. It is convenient to take the NLSE as $H_n^{(2)}$, and use it to find a second nonlinear system with Hamiltonian $H^{(1)}_n$ and additional level $E_0$. 

The above procedure is implemented as follows. We require
\begin{equation}
H_n^{(2)}=-\frac{d^2}{dx^2}-\kappa|\psi_n|^2=\hat{A}_n\hat{A}^{\dagger}_n+E_0,
\end{equation}
and solve
\begin{equation}\label{eq:ric}
W^2_n(x)+W'_n(x)=-\kappa|\psi_n|^2-E_0
\end{equation}
to find the level-dependent superpotential, $W_n(x)$. The choice of $E_0$ is arbitrary, as long as it is lower than the ground state energy of the second Hamiltonian, $E_1$. We can thus engineer the available solutions by making an informed choice for $E_0$. Our first Hamiltonian has the level dependent form
\begin{equation}\label{eq:H1n}
H_n^{(1)}=-\frac{d^2}{dx^2}+W^2_n(x)-W'_n(x)+E_0.
\end{equation}
Using the superpotential obtained from Eq. (\ref{eq:ric}), we can find the eigenfunctions from the condition $\phi_n=(E_n-E_0)^{-1/2}(-d/dx+W_n(x))\psi_n$. Furthermore, we can obtain a definition of the superpotential in terms of the two eigenfunctions:
\begin{equation}
W_n(x)=\frac{(E_n-E_0)^{1/2}\psi_n-\phi'_n}{\phi_n},
\end{equation}
where the prime indicates a derivative in $x$.

\begin{figure}[h]
\centering
\includegraphics[width=9cm]{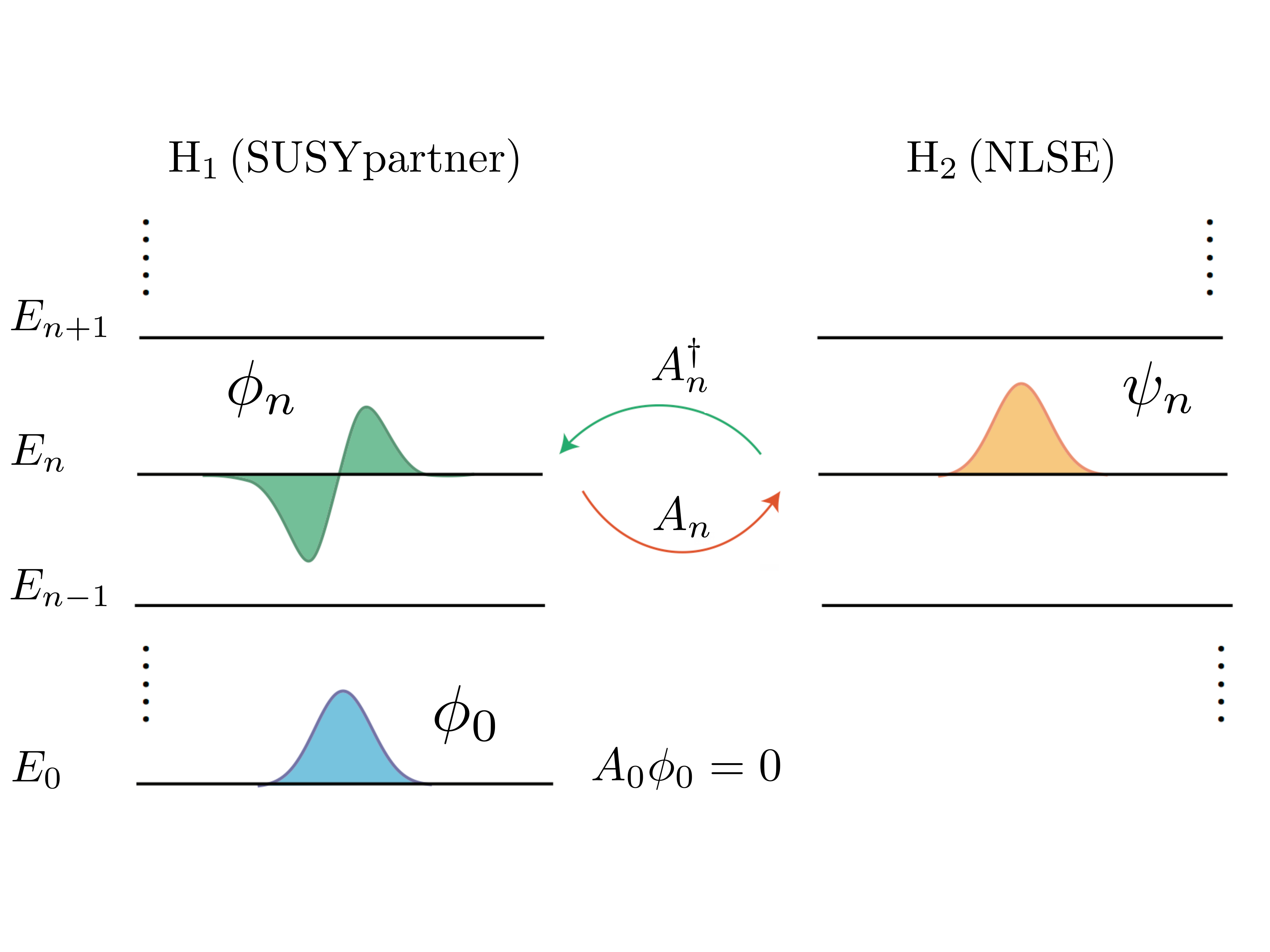}
\caption{A visualisation of the nonlinear SUSY-QM process for our example solution. As not all solutions to the NLSE are known, we simply label the soliton solution we chose as `$E_n$', the `nth' solution. The diagram is set out in levels, in analogy with linear SUSY-QM (e.g. see \cite{SUSYQM}, page 18), however, here each `level' $E_n$ represents a distinct solution of the nonlinear SUSY-QM problem. $\phi_0$ represents a solution of the NLSE's SUSY-partner corresponding to energy $E_0$, such that no counterpart solution of the NLSE exists. The equation it solves appears to be a scale-free nonlinear equation, but it can be easily transformed into a linear equation, as detailed in the main text.}
\label{fig1}
\end{figure}

This allows us to find the SUSY-QM partner to the NLSE purely in terms of the eigenfunctions of the two equations; we essentially have a coupled nonlinear system. There are various ways to write Eq. (\ref{eq:H1n}) such that the superpotential is eliminated, however too much substitution between $\psi_n$ and $\phi_n$ with the goal to eliminate $\psi_n$ terms will lead to the tautology $E_n\phi_n=E_n\phi_n$. Our preferred form of the nonlinear equation corresponding to $H_n^{(1)}\phi_n$, is thus
\begin{widetext}
\begin{equation}\label{eq:susy1}
\phi''_n-2\frac{(\phi_n')^2}{\phi_n} -\kappa\,\phi_n|\psi_n|^2+2(E_n-E_0)^{1/2}(\psi_n\frac{\phi'_n}{\phi_n}-\psi_n')=E_n\,\phi_n.
\end{equation}
\end{widetext}
This can alternatively be written as:
\begin{widetext}
\begin{equation}
\frac{d^{2}}{dx^{2}} \phi_n-8\,(\frac{d}{dx} \sqrt{\phi_n})^2 -\kappa\,\phi_n|\psi_n|^2-2\sqrt{\Delta E_n}\,\frac{d}{dx}\bigg(\frac{\psi_n}{\phi_n}\bigg)=E_n\,\phi_n, \label{susy2}
\end{equation}
\end{widetext}
where $\Delta E_n=E_n-E_0$. We can think of the system described by Eq. (\ref{susy2}) as supersymmetric - in the quantum mechanical sense - to the NLSE, Eq. (\ref{eq:sch}). The concept of `energy levels' may seem unnatural in the context of nonlinear equations, but here, we can think of the functions $\psi_n$ as distinct eigenfunctions of the NLSE, $\phi_n$ as distinct eigenfunctions of its SUSY partner, and $\phi_0$ as an eigenfunction corresponding to energy $E_0$, for which no corresponding eigenstate of the NLSE exists. Excluding $E_0$, the two nonlinear systems will share an identical spectrum. It is clear that a function, $\psi_n$, which allows this equation to be solved for a given $\phi_n$, will solve the NLSE. The additional energy dependent term comes from the derivative of the superpotential, and although unusual, cannot be avoided. Figure \ref{fig1} shows a graphical way to visualise our nonlinear SUSY transformation scheme.

It is now possible to generalise the above result. It should be clear that for any nonlinear equation which can be written in the form:
\begin{equation}
-\psi''_n-N(\psi_n) \psi_n=E_n \psi_n, \label{NL}
\end{equation}
where $N(\psi_n)$ represents a nonlinear operator, there exists a SUSY-QM partner equation which will have the form,
\begin{widetext}
\begin{equation}
\frac{d^{2}}{dx^{2}} \phi_n-8\,(\frac{d}{dx} \sqrt{\phi_n})^2 -N(\psi_n)\,\phi_n-2\sqrt{\Delta E_n}\,\frac{d}{dx}\bigg(\frac{\psi_n}{\phi_n}\bigg)=E_n\,\phi_n, \label{susy3}
\end{equation}
\end{widetext}
given one can solve the Riccati equation, Eq. (\ref{eq:ric}). This equation can be greatly simplified by making the substitution $\phi_n=1/u_n$, and rearranging:
\begin{widetext}
\begin{equation}
- u''_n -N(\psi_n)\,u_n-2\sqrt{\Delta E_n}\,(u'_n u_n \psi_n+\psi'_n u_n^2)=E_n\,u_n. \label{NLSUSYeq}
\end{equation}
\end{widetext}

Equation (\ref{NLSUSYeq}) now maintains a more typical format, and is also clearly still nonlinear. It is important to observe that the nonlinear terms of this equation do not arise from the $N(\psi_n)$ term, but from the requirement that the superpotential, $W_n$, is forced to be eigenfunction-dependent. It may seem that one can set the nonlinear operator in Eq. (\ref{NLSUSYeq}) to zero, and receive a nonlinear equation partnered to an equation for a free particle; this is incorrect, as in this case the superpotential is clearly {\em eigenfunction-independent}. The $\psi_n$ in Eq. (\ref{NLSUSYeq}) must belong to a nonlinear equation, in order for our SUSY construction to be valid. 

We shall now show that when a solution to Eq. (\ref{NL}) ceases to exist, Equation (\ref{NLSUSYeq}) reduces to the \schr equation for a free particle. 

\section{The `vacuum' equation}
From linear SUSY-QM, we have the condition $\hat{A}\phi_0=0$ (annihilation of the ground state), which gives us the simple relation $W(x)=-d/dx \ln{\phi_0}=-\phi_0'/\phi_0$. For us, this condition is no longer universal, and only valid when $\phi_n \rightarrow \phi_0$ and $\psi_n \rightarrow 0$. Making the standard quantum-mechanical substitution $E_0\rightarrow i(d/dt)$,  Eq. (\ref{susy3}) reduces to:
\begin{equation}\label{eq:sf}
i\dot{\phi_0}-\phi_0''+2\frac{(\phi_0')^2}{\phi_0}=0,
\end{equation}
where the dot indicates a time derivative. This equation can be easily solved, and its bound state solutions are hyperbolic secants. Equation (\ref{eq:sf}) can be written as the Lax equation seen in the work of Zaharov and Shabat \cite{ZS},
\begin{equation}\label{eq:Lax}
i\frac{\partial \hat{L}}{\partial t}+[\hat{L},\hat{M}]=0,
\end{equation}
for the following Lax pair:
\begin{equation}
\hat{L}=\frac{d}{dx}+\frac{1}{\phi_0},
\end{equation}
and
\begin{equation}
\hat{M}=\frac{d}{dx}+\frac{\phi_0'}{\phi_0^2}+\frac{1}{\phi_0}.
\end{equation}

Interestingly, Eq. (\ref{eq:sf}) is reminiscent of the equation for the propagation of an optical field seen in what is known as scale-free optics \cite{SF}, although with some important differences in the dimensionality and the use of intensities instead of the ratio of envelope fields. The system is known as `scale-free' due to the fact it is intensity independent. This property is mirrored in our equation by the fact that the amplitude of the solutions play no role in the dynamics. This is an indication that the equation is in fact a linear equation in disguise, and on making our earlier substitution $\phi_0=1/u_0$, Eq. (\ref{eq:sf}) essentially becomes the \schr equation for a free particle, and it becomes clear that the Lax pair above is 'fake'; it says nothing about the integrability of the system \cite{fakelp}.


\section{An example solution}

\begin{figure}[h]
\centering
\includegraphics[width=10cm]{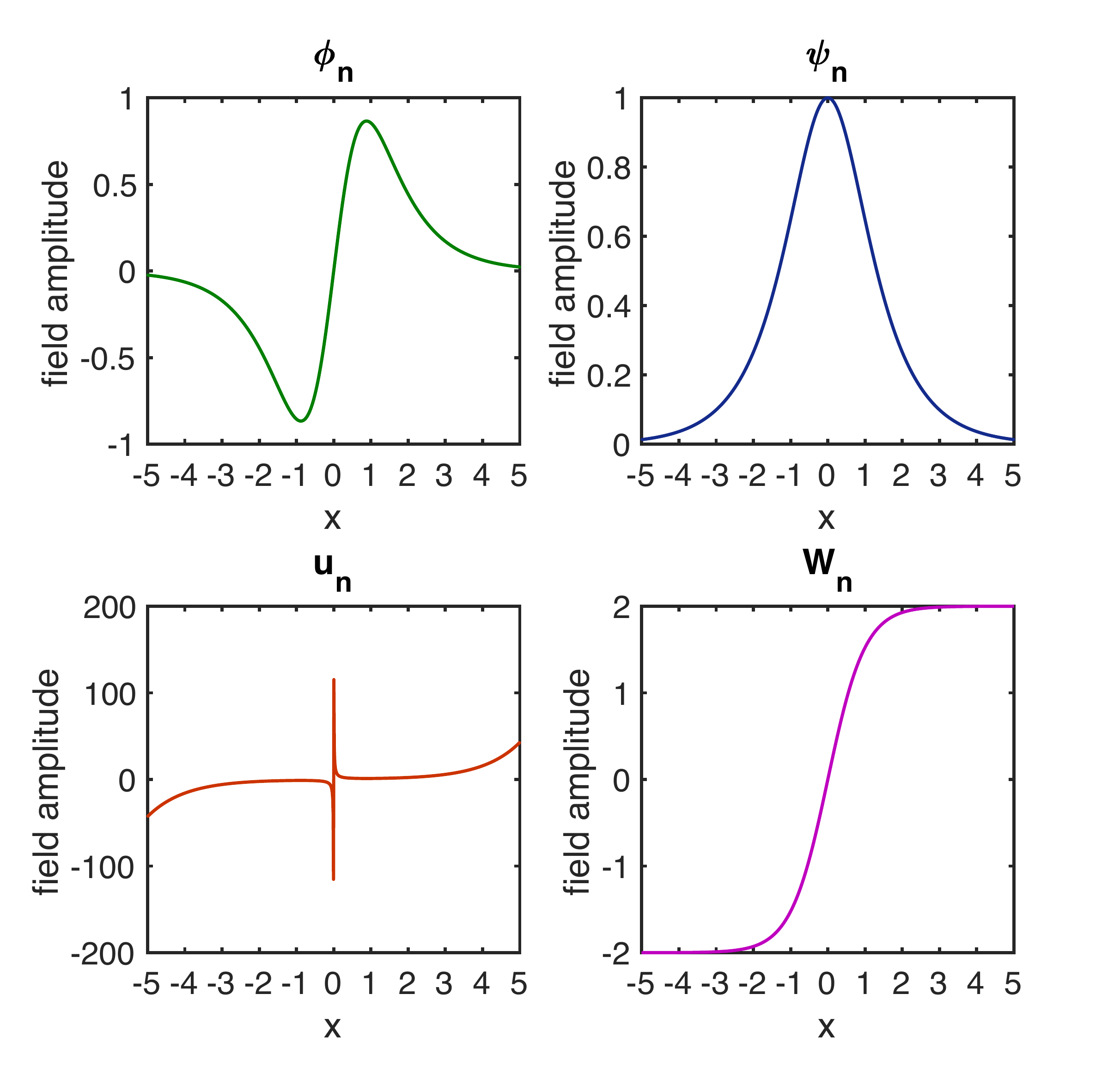}
\caption{Example SUSY functions for a particular solution of the NLSE (the soliton), plotted on the $x$ axis in units of $\hbar=2m=1$. a) an eigenfunction of Eq. (\ref{susy2}), $\phi_n(x,t)=\sqrt{3}\,\sech{x}\tanh{(x)} e^{(-i E_n t)}$, b) the corresponding eigenfunction of the NLSE, $\psi_n(x,t)=\sech{ x} e^{(-i E_1 t)}$, c) the corresponding eigenfunction of Eq. (\ref{NLSUSYeq}), $u_n(x,t)=\frac{1}{\sqrt{3}}cosh{(x)}\coth{(x)}e^{(i E_n t)}$ and d) the superpotential $W_n=2\tanh{(x)}$ which connects $\phi_n$ and $\psi_n$.}
\label{fig2}
\end{figure}

For solutions of the NLSE (with $N(\psi_n)=-2|\psi_n|^2$) of the form $\psi_n=\sech{x} e^{-i E_n t}$, Eq. (\ref{eq:ric}) can be solved and the particular solution $W_n=2 \tanh{(x)}$ can be found, given the choice of the parameter $E_0=-4$. Note that any value of $E_0<-1$ can be chosen without `breaking' SUSY \cite{SUSYQM}. We can find $\phi_n = \sqrt{3}\sech{x}\tanh{(x)}e^{-i E_n t}$ from the definitions of $A_n^{\dagger}$ and $\psi_n$. From here, we can find a solution to Eq. (\ref{NLSUSYeq}) of the form $u_n=\frac{1}{\sqrt{3}}\coth{(x)}\cosh{(x)}e^{i E_n t}$. Eq. (\ref{susy3}) may be preferable to some, as its solution is clearly a soliton, whereas the solution to Eq. (\ref{NLSUSYeq}) is, in this case, singular. This process is visualised in a form analogous to the level diagrams seen in traditional SUSY-QM in Figure \ref{fig1}. The exact form of the functions detailed above can be seen in Figure \ref{fig2}. 

\section{Conclusions} 
We have presented a new method for constructing nonlinear evolution equations by extending the formalism of SUSY-QM to nonlinear systems. The result is an equation dependent on both the eigenstates of the original nonlinear equation and the eigenstates of the new equation. When the eigenstates of the original nonlinear equation vanish, the new equation can be reduced to a linear one using a simple transformation. Our scheme is easily extended to all NLSE-type evolution equations containing second order derivatives, and can be used to obtain SUSY-partner equations for a large variety of nonlinear models, establishing a web of previously unknown `dualities' between soliton solutions and their respective equations.

\section*{Acknowledgements} F.B. would like to acknowledge funding from the German Max Planck Society for the Advancement of Science (MPG) and  the International Max Planck Partnership (IMPP) between MPG and the Scottish SUPA Universities. R.H. acknowledges funding from the Condensed Matter Centre for Doctoral Training (CM-CDT). We also thank Takayuki Tsuchida for useful remarks.

\end{document}